\documentclass[12pt]{iopart}

\usepackage{graphicx} 
\newcommand\ddfrac[2]{\frac{\displaystyle #1}{\displaystyle #2}}

\begin{document}

\title[]{Quantum Transport through Asymmetrical Molecular channel Azulene: Role of Orbital Interference }

\author{Koushik R. Das and Sudipta Dutta*}

\address{Department of Physics, Indian Institute of Science Education and Research (IISER) Tirupati, Tirupati - 517619, Andhra Pradesh, India.}
\ead{sdutta@iisertirupati.ac.in}
\vspace{10pt}
\begin{indented}
\item[]December 2024
\end{indented}

\begin{abstract}
   We investigate electron transport through azulene molecule with four distinct electrode contact geometries using the non-equilibrium Green's function formalism within the tight-binding Hamiltonian. Employing the Q-matrix approach, we analyze quantum interference (QI) among the molecular orbitals in each contact configuration. Our results reveal distinct transmission profiles and varying current responses among configurations, with configuration 1-3 displaying the highest conductivity at higher bias due to strong constructive interference of the Highest Occupied Molecular Orbital (HOMO). Conversely, configuration 5-7 exhibit weak conductance and antiresonance at the Fermi energy, attributed to dominant destructive interference among the frontier molecular orbitals. Configuration 2-6 is found to exhibit asymmetric I-V characteristics, due to the dipolar nature of the azulene molecule. These findings underscore the significance of QI effects in shaping the transport properties of azulene, and molecule-based devices in general.
\end{abstract}

%
%
%
%
%

\section{Introduction}

Electron transport in molecular systems stands at the forefront of nanotechnology, offering promising prospects for the development of advanced nanoelectronic devices\cite{joachim2000electronics,huang2020recent,gupta2023nanoscale,li2023molecular}. In recent years, remarkable advancements have been made in both experimental techniques and theoretical studies  aimed at unraveling the intricacies of electron transport phenomena\cite{metzger2018quo,homma2023dependence,datta2000nanoscale}. Among the diverse array of molecules investigated for their electron transport properties, conjugated systems have garnered considerable attention due to the presence of delocalized electrons, which serve as primary carriers for conduction.

Benzene and naphthalene are some of the molecules that are extensively studied, both theoretically\cite{choi2005role,yan2007ab} and experimentally\cite{liu2015highly}. These molecules, characterized by their alternant hydrocarbon structure \cite{coulson1940note} and aromatic nature with a zero dipole moment, have provided valuable insights into fundamental electron transport mechanisms. However, the exploration of non-alternant molecules introduces a new dimension to this research landscape. Azulene, a non-alternant isomer of naphthalene, features a unique structural asymmetry, with fused five- and seven-membered rings\cite{bergmann1971aromaticity}. This structural asymmetry causes the five-membered ring to act as ``acceptor" and the seven membered ring as ``donor" to satisfy the H\"uckel's rule for aromaticity, thereby gaining a net dipole moment\cite{tobler1965microwave}. This renders the electron transport through azulene distinct from naphthalene, as has been shown in published literature \cite{dutta2008comparative,treboux1998asymmetric,el2016first,alqahtani2023influence}.

In this study, we explore the effect of quantum interference (QI) among molecular orbitals\cite{tsuji2017frontier} and its relation to the current conductivity of the molecule contacted at different atomic sites. Our objective is to elucidate the underlying mechanisms responsible for the site-dependent asymmetric current response observed in such systems. Importantly, the insights gained from our analysis are expected to have broader implications, irrespective of the distinction between alternant and non-alternant systems, and offering a generic framework for understanding electron transport in molecular architectures.

\section{Model and Method}
The device consists of azulene molecule attached in between two metal electrodes, namely, L and R, as depicted in Fig.1. All couplings are taken to be through the bonds. We assume the electrodes couple only with the nearest sites of the molecule, termed as contact sites and are denoted by numerics in Fig.1. 

\begin{figure}[h]\centering
	\includegraphics[width=0.5\textwidth]{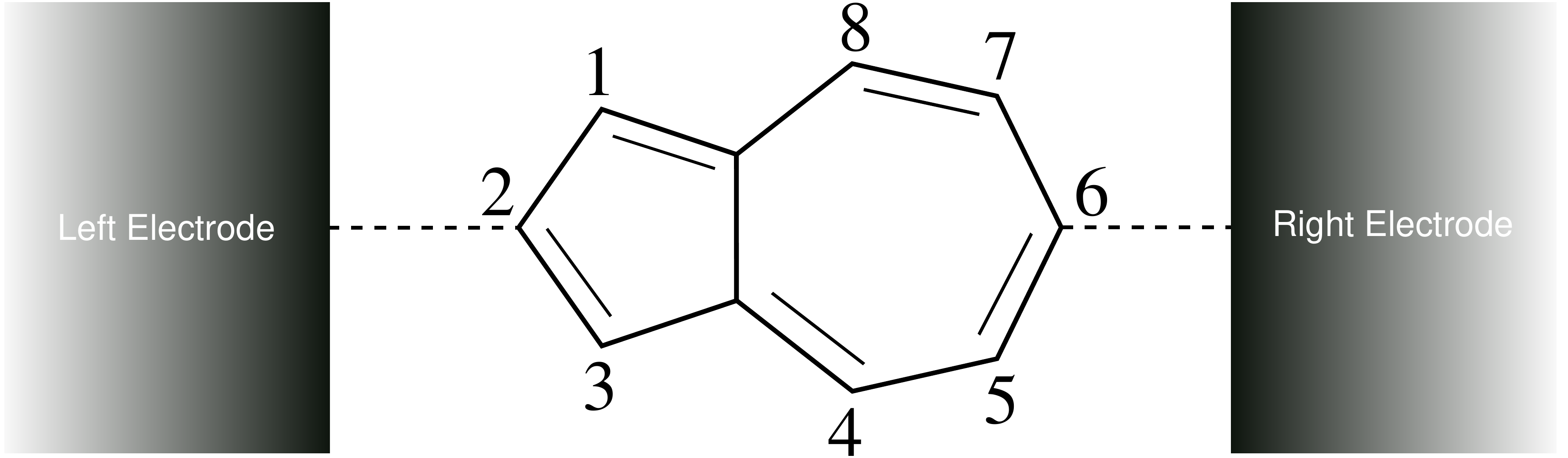}
	\caption{Schematic representation of an azulene molecule sandwiched between a pair of electrodes in the 2-6 contact geometry configuration. The numeric values indicate the other possible contact sites.}
\end{figure}

The device has been modeled within tight-binding (TB) Hamiltonian as follows:

\begin{equation}
H_{mol} = \sum_{<ij>}^{N} t(c_i^\dagger c_j+c_j^\dagger c_i) - \sum_{i}^{N}V_i c_i^\dagger c_i
\end{equation}

\noindent where, $t$ is the hopping parameter among sites and $V_i$ is the potential at the $i^{th}$ site due to external applied bias. $c_i$ ($c_i^\dagger$) are the annihilation (creation) operator and $N$ is the total number of atoms (sites) in the molecule. The hopping is considered only up to the nearest neighbor. Since we are dealing with electron transport, the potential energy due to the applied bias is $-V_i$. The potential distribution between the left ($L$) and right ($R$) electrodes is given by\cite{das2024asymmetric}:

\begin{equation}
V_i = - \ddfrac{V_L r_{iL} + V_R r_{iR}}{r_{iL}+r_{iR}} 
\end{equation}

We implement the non-equilibrium Green's function (NEGF) formalism\cite{datta1997electronic} to obtain the transmission function. It is defined as the total probability of an electron with an energy $E$ to traverse between electrodes through the device region and it is estimated as follows:

\begin{equation}
T(E,V)=Tr[\Gamma_L G(E,V)\Gamma_R G^\dagger(E,V)]
\end{equation}

In the above expression, $G(E,V)$ is the retarded Green's function. $\Gamma_{L/R}$ is the spectral density that results in energy level broadening due to $L/R$ electrode and it is defined as 
\begin{equation}
\Gamma_{L/R}=i~ (\Sigma_{L/R}-\Sigma^\dagger_{L/R})
\end{equation}	

\noindent where $\Sigma_{L/R}$ are the self-energies for $L/R$ electrodes, that describe the molecule-electrode coupling. 

The device Green's function, $G(E,V)$ is defined as, 
\begin{equation}
G(E,V)=((E+i\eta)I-H)^{-1} = ((E+i\eta)I-H_{mol}-\Sigma_L-\Sigma_R)^{-1}
\end{equation}

\noindent where an infinitesimal number $\eta$ is used to avoid the zero division. The current through the device is then calculated as weighted average over the transmission function, $T(E,V)$, as given by the Landauer-B\"{u}ttiker formula\cite{datta1997electronic,datta2000nanoscale}:
\begin{equation}
I(V)= {\frac{2e^{2}}{h}} \int_{\mu_{L}}^{\mu_{R}} T(E,V) dE 
\end{equation}

\noindent where $e$ is the electronic charge, $h$ is the Planck's constant. $\large\mu\small_{L/R}$ are the chemical potentials of the $L/R$ electrodes, defined as $\large{\mu}\small_{L/R} = E_F \mp eV/2 $, with $\large\mu\small_R - \large\mu\small_L=eV$; $E_F$ being the Fermi energy of the electrode at zero-bias. The interval $[\mu_L(V),\mu_R(V)]$ i.e $[-V/2,V/2]$ denotes the energy region that contributes to the current integral and is denoted as the bias window. This is in accordance with the fact that electrons near the $E_F$ will only contribute to the total current.

We apply this approach in conjunction with the NEGF to analyze the calculated transmission behavior for various systems. In this work, we solve the TB Hamiltonian, setting $t$ as the unit, with molecule-electrode coupling in the weak coupling limit ($\Sigma_{L/R}=0.05$) and $E_F=0$. External bias is varied from $-1V$ to $1V$ with $0.1V$ intervals and corresponding current is calculated.

\section{Results and Discussion}   

We investigate the variation of conductance for varying contact sites of the azulene molecule. Configuration 2-6 features electrodes attached to sites located across both the five- and seven-membered rings along the symmetry line (Fig.1). Conversely, configuration 1-3 has both electrode connections situated solely on the five-membered ring, while configurations 4-7 and 5-7 have both connections exclusively on the seven-membered ring. Such contact geometries are considered to investigate the quantum interference effect among the transmitted electrons in azulene, in terms of the wavefunction phase that is embedded in the Green's function description. 

Examination of the zero-bias transmission functions, as depicted in Fig.2(a), reveals a close qualitative agreement with GW-based calculations\cite{xia2014breakdown}. Due to the non-alternant nature of azulene, the absence of electron-hole symmetry leads to asymmetric transmission functions below and above $E_F$. Notably, the 5-7 configuration exhibits a prominent antiresonance dip at $E_F$, distinguishing it from the other configurations where non-zero transmission persists in the off-resonance regime.

Our focus lies within the region $[-0.5t,0.5t]$, since the current calculations are done within this energy window. The highest occupied molecular orbital (HOMO) and the lowest unoccupied molecular orbital (LUMO) situate at $-0.45t$ and $0.4t$, which are within this energy window. It appears that these frontier orbitals are acting as the conducting channels for the electrons and contributing towards the total current flowing through the molecule. 

\begin{figure}[h]\centering
	\includegraphics[width=0.5\textwidth]{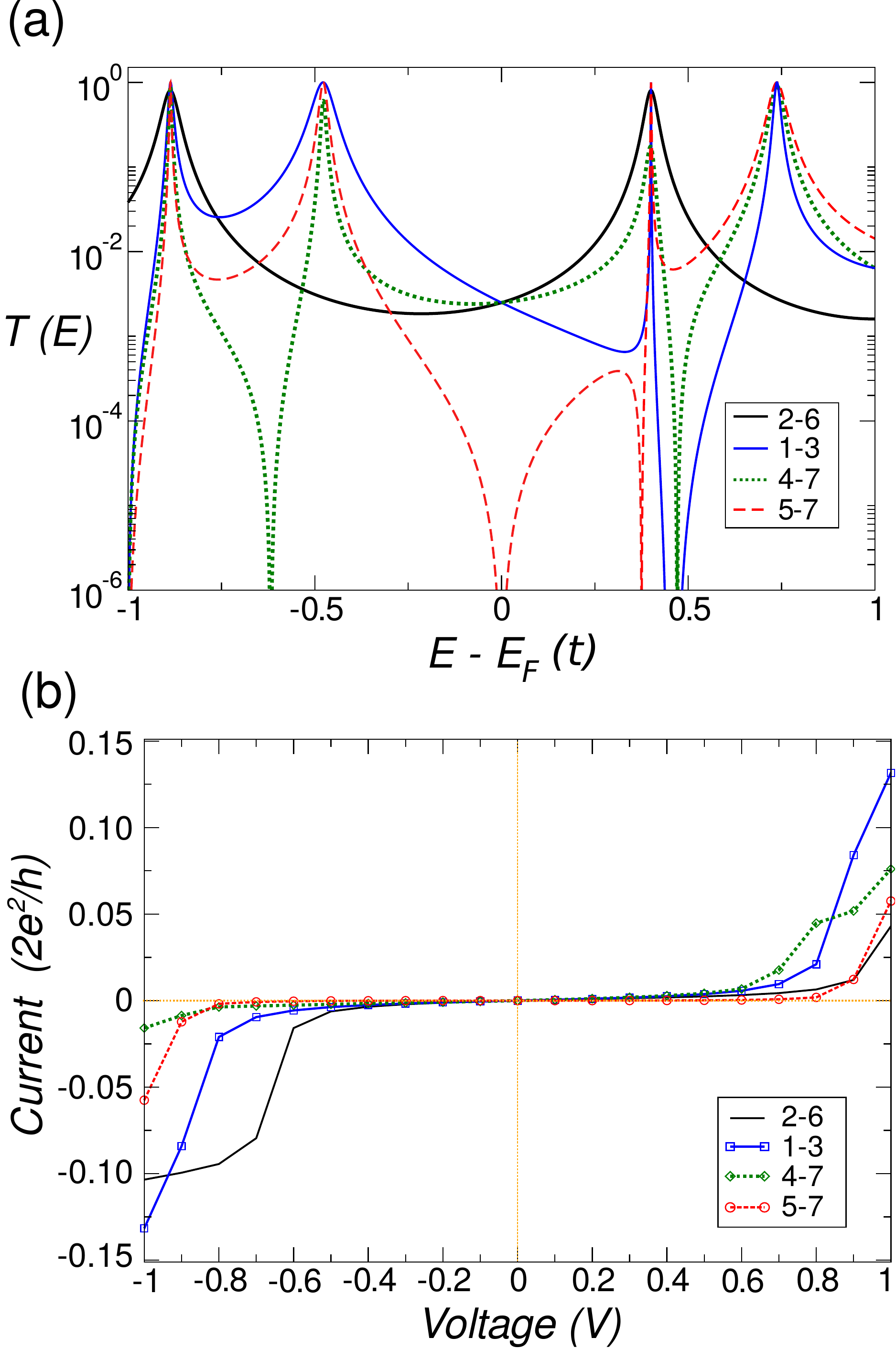}
	\caption{(a) Zero-bias normalized transmission functions and (b) I-V characteristics of azulene with varying contact geometries.}
\end{figure}

As can be seen, the configuration 5-7 has the lowest current response. Notably, configuration 4-7 displays dissimilar $I-V$ characteristics in positive and negative bias regimes, attributable to the asymmetrical nature of the configuration when attached between the electrodes. An interesting observation arises from configuration 2-6, which demonstrates significantly higher current in the negative regime compared to the positive regime, revealing its rectifying nature. This behavior can be attributed to the non-zero dipole moment of azulene\cite{tobler1965microwave}. The five- and seven-membered rings together form a ``acceptor-donor" system as per the H\"uckel's aromaticity rule, with the former having a higher electric potential energy than the latter.

According to our convention, forward/positive bias regime is defined as the left-electrode at positive potential and the right-electrode at negative potential, resulting in current flowing from left to right, and vice-versa for electron transfer. In the case of configuration 2-6, under application of forward bias, electrons injected from the right-electrode into the seven-membered ring (donor) require a sufficiently large injection energy to overcome the potential energy barrier between the ``donor" and ``acceptor" and traverse the five-membered ring to reach the left-electrode, thereby limiting the overall forward-bias current. Conversely, under reverse bias, electrons injected from the left-electrode into the five-membered ring move successively to the seven-membered ring and reach the right-electrode without encountering such potential energy barrier, resulting in a higher current output compared to the forward bias regime. This phenomenon has also been observed in previous literature\cite{dutta2008comparative,treboux1998asymmetric}. 

Configuration 1-3 also exhibits a higher current response than 4-7 and 5-7, attributable to the larger area under the corresponding transmission function curve (Fig.2(a)), particularly below $E_F$. These calculations highlight that different configurations, even with electrode attachments on the same ring, such as in the case of configuration 4-7 and 5-7, can manifest  drastically different current responses.

Note that, the sudden fluctuations in transmission values in the range of  $0.4t - 0.5t$ for 1-3, 4-7 and 5-7 configurations appear to be the signature of the Fano resonances, which can arise due to successive constructive and destructive interferences between the molecular orbitals (MOs) within a very narrow energy range.

These transmission features discussed above exhibit a direct one-to-one relation with the interferences among molecular orbitals (MOs), and can be further probed and visualized using the Q-matrix formalism, as proposed by Gunasekaran et.al.\cite{gunasekaran2020visualizing}. Within the NEGF formalism, the Green's function is generally described in the atomic orbital (AO) basis. However, to extract QI information, a basis transformation of the Green’s function from the AO to the MO basis is necessary. This requires a transformation matrix $P$ whose columns are comprised of the eigenvectors of $G$ such as $P^{-1}GP$ is a diagonal matrix. Within this MO basis, the transmission function $T$ becomes a sum over all MOs, i.e incorporating both the non-interfering diagonal terms $T_i$ and interfering off-diagonal terms $T_{ij}$ explicitly,	

\begin{equation}
T=\sum_{i}T_i + \sum_{i>j} T_{ij}
\end{equation}

This is obtained by evaluating the Q-matrix, defined as,

\begin{equation}
Q(E)=(P^\dagger \Gamma_L G P)\circ(P^{-1} \Gamma_R G^\dagger P^{-1 \dagger})^T
\end{equation}

Here, [$\circ$] denotes the entrywise (Schur) product, and $[*]^T$ is matrix transpose. It can be shown that,

\begin{equation}
T=\sum_{ij}Q_{ij}
\end{equation}

\begin{figure}[h]\centering
	\includegraphics[width=1.0\textwidth]{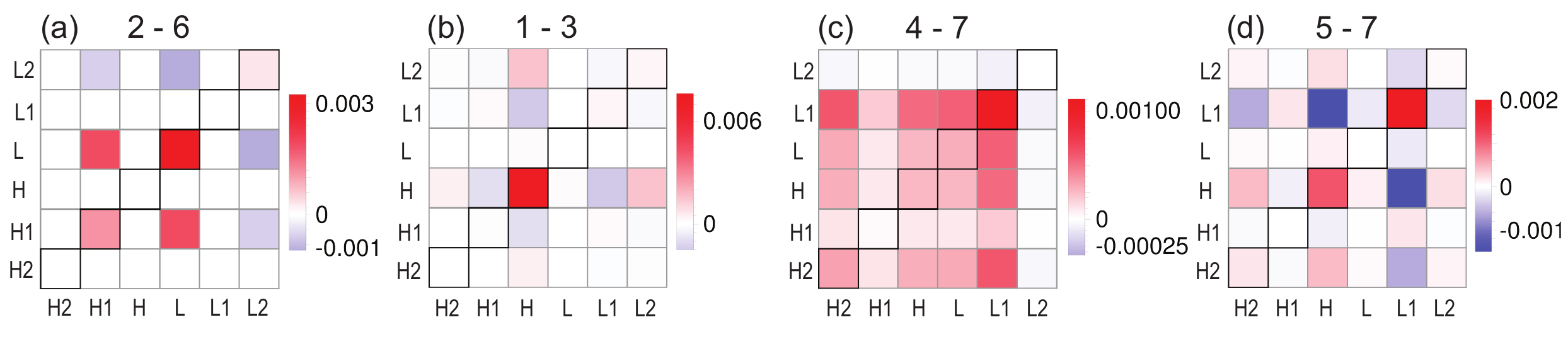}
	\caption{Zero-bias color maps of the Q-matrices at Fermi energy for (a) 2-6 (b) 1-3 (c) 4-7 and (d) 5-7 configurations, showing interferences among the frontier MOs. Note that the color scale is normalized to $Q_{max}$ for each configuration. The H2, H1, H, L1, L2 indicate HOMO-2, HOMO-1, HOMO, LUMO, LUMO+1, LUMO+2, respectively.}
\end{figure}

The above expression indicates that the transmission function $T$ and the Q-matrix are equivalent to each other. To investigate the QI effect we calculate the Q-matrices for specific energy and bias for each configuration and present them as color-maps. In Fig.3, we present the Q-matrices calculated at Fermi energy for zero bias. In these maps, red and blue colors represent constructive and destructive interferences, respectively, with the intensity of the color indicating the magnitude of the interference. Note that the diagonal is aligned along top-right to bottom-left.

The LUMO dominated constructive interference in 2-6 (Fig.3(a)) and HOMO dominated constructive interference in 1-3 (Fig.3(b)) configurations lead to high transmission and subsequent higher current response. It can be noted from Fig.3(c) that the configuration 4-7 displays constructive interferences among multiple frontier MOs. This may initially lead to the expectation that the configuration 4-7 to have the highest transmission at the Fermi energy due to these combined effects, however, the magnitude of such interferences (as can be noted from the color scale beside the respective plots in Fig.3) is lower as compared to that of the configurations 2-6 and 1-3, resulting in comparable transmission and current output. Unlike the rest of the configurations, the 5-7 configuration exhibits strong destructive interference between the HOMO and the LUMO+1, nullifying their individual contributions and giving rise to the transmission antiresonance observed in Fig.2(a).  

\begin{figure}[h]\centering
	\includegraphics[width=0.5\textwidth]{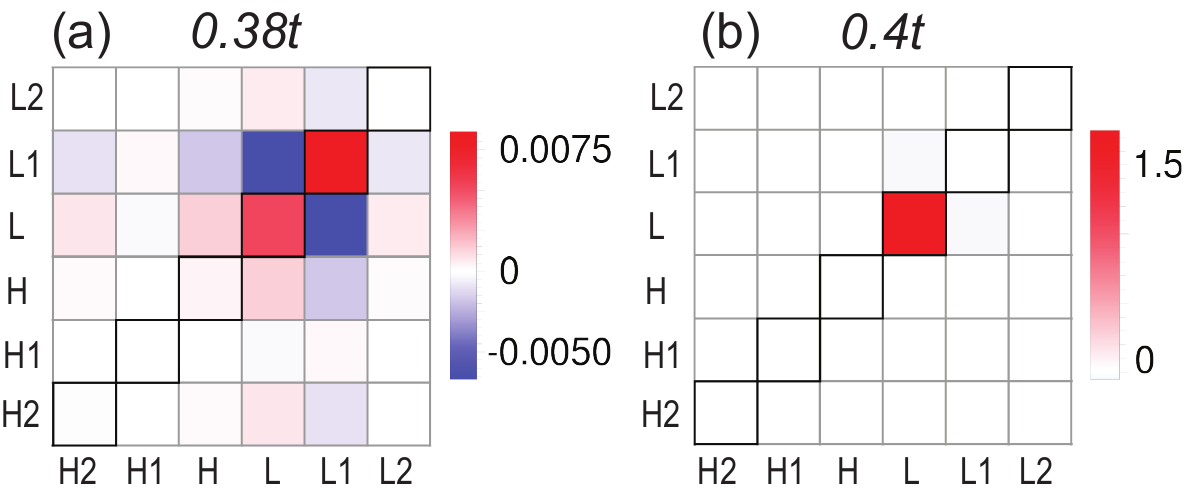}
	\caption{Zero-bias color maps of the Q-matrices at energy values (a) $0.38t$ and (b) $0.4t$ for the 5-7 configurations, showing consecutive destructive and constructive interferences in a narrow energy window that leads to Fano resonance in the transmission plot. Note that the color scale is normalized to $Q_{max}$ for each configuration.}
\end{figure}

Furthermore, to confirm the appearance of the Fano resonance in the range of $0.4t - 0.5t$ in the transmission plot (Fig.2(a)), we calculate the Q-matrices at corresponding energies. In Fig.4, we present the same at the transmission dip ($0.38t$) and peak ($0.4t$) positions for the 5-7 configurations. As can be seen, the destructive interference between the LUMO and LUMO+1 at $0.38t$ creates the dip in the transmission, immediately followed by the transmission peak at $0.4t$ arising from the LUMO. Therefore, by changing the contact geometry, one can induce such Fano resonance which in turn provides control to regulate the current response.

These observations underscore the importance of analyzing  the corresponding transmission functions and interpreting in context of quantum interference among MOs. The arguments and methodology presented here highlight the intricate relationship between the molecular structure, orbital interactions and electron transport dynamics, are general in nature and can provide valuable insight into the electronic transport properties of molecular systems.

\section{Conclusion}	
In this study, we investigate the electron transport through the azulene molecule in four specific contact geometry configurations using the NEGF formalism within the tight-binding model. The Q-matrix approach is employed to understand quantum interference among the MOs of each configuration. The non-alternant nature of azulene results in asymmetric transmission functions across the Fermi energy due to the absence of electron-hole symmetry. Each configuration exhibits distinct transmission profiles, leading to varying current responses. Notably, configuration 1-3, having both electrodes connected to the five-membered ring, and configurations 4-7 and 5-7, with both electrode connections on the seven-membered ring, display drastically different current responses. This variation is attributed to the degree of interference among the MOs. The dominant destructive interference in 5-7 results in a transmission antiresonance at the Fermi energy, leading to weak conductance. However, the strong contribution from the HOMO in 1-3 renders it the most conducting. We also show the appearance of Fano resonance in such molecular transport channels by varying the contact geometry that causes strong destructive and constructive interferences among the MOs within a narrow energy window. These findings highlight the crucial role of QI effects in determining the transport properties of molecular systems, offering insights for the design of electronic devices.

\section{Data Availability Statement}
The data volume is very large. The data that support the findings of this study are available upon reasonable request from the authors.

\section{Acknowledgement}
K.R.D and S.D. thank IISER Tirupati for intramural funding and the Science and Engineering Research Board (SERB), Department of Science and Technology (DST), Government of India for research grant CRG/2021/001731.

\section{References}
\bibliographystyle{iopart-num}

\providecommand{\newblock}{}

\end{document}